\documentclass[amsmath,amssymb,
 aps,twocolumn,pra]{revtex4-1}
\usepackage{graphicx}
\usepackage{dcolumn}
\usepackage{textcomp}
\begin{document}
\title{Dispersion tailoring in wedge microcavities for Kerr comb generation}

% \author[1]{L. Fujii}
% \author[1]{M. Inga}
% \author[1]{J. H. Soares}
% \author[1]{Y. A. V. Espinel}
% \author[1]{T. P. Mayer Alegre}
% \author[1,*]{G. S. Wiederhecker}
\author{L. Fujii}
\author{M. Inga}
\author{J. H. Soares}
\author{Y. A. V. Espinel}
\author{T. P. Mayer Alegre}
\author{G. S. Wiederhecker}
\email{Corresponding author: gsw@unicamp.br}
\affiliation{Photonics Research Center, Applied Physics Department, Gleb Wataghin Physics Institute, P.O. Box 6165, University of Campinas - UNICAMP, 13083-970 Campinas, SP, Brazil}
% \affil[1]{Photonics Research Center, Applied Physics Department, Gleb Wataghin Physics Institute, P.O. Box 6165, University of Campinas - UNICAMP, 13083-970 Campinas, SP, Brazil}

% \affil[*]{Corresponding author: gsw@unicamp.br}

%% To be edited by editor
% \dates{Compiled \today}

% \ociscodes{(140.3490) Lasers, distributed feedback; (060.2420) Fibers, polarization-maintaining;(060.3735) Fiber Bragg gratings.}

%% To be edited by editor
% \doi{\url{http://dx.doi.org/10.1364/XX.XX.XXXXXX}}

\begin{abstract}
The shaping of group velocity dispersion in microresonators is an important component in the generation of wideband optical frequency combs. Small resonators -- with tight bending radii -- offer the large free-spectral range desirable for wide comb formation. However, the tighter bending usually limit comb formation as it enhances normal group velocity dispersion. We experimentally demonstrate that engineering the sidewall angle of small-radius ($\sim 100$~$\mu$m), 3~$\mu$m-thick silica wedge microdisks enables dispersion tuning in both normal and anomalous regimes, without significantly affecting  the free spectral range.  A microdisk with wedge angle of $55^{\circ}$ (anomalous dispersion) is used to demonstrate a 300~nm bandwidth Kerr optical frequency comb.
\end{abstract}

% \setboolean{displaycopyright}{true}

\maketitle

%\section{Introduction}

\begin{figure*}[ht]
\centering
\includegraphics[scale=1]{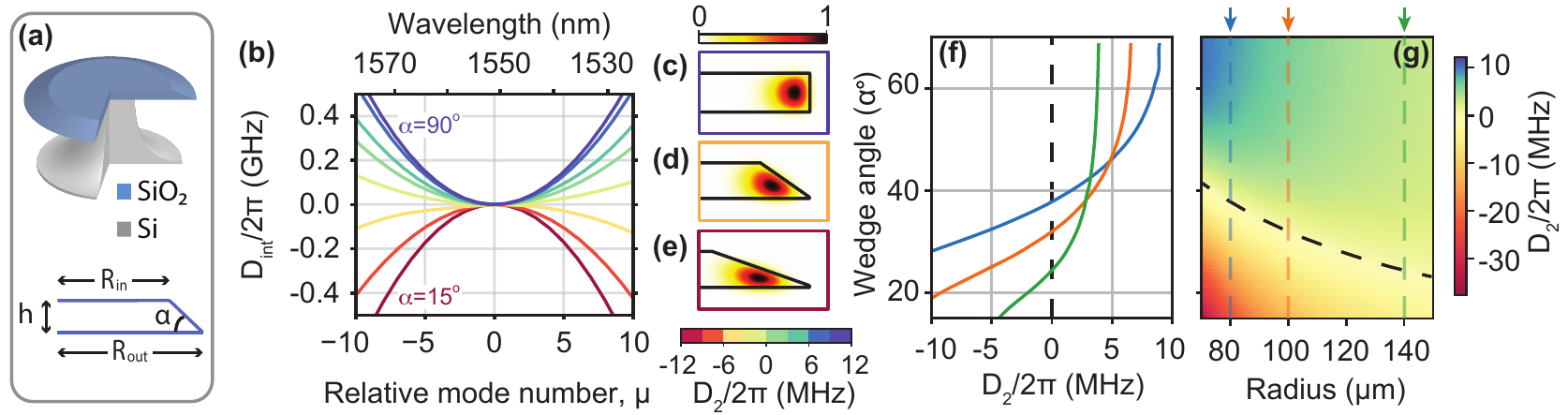}
\caption{a) Silica disk schematic. b) Simulated dispersion curves for $R_\text{out}=100~\mu$m, $h=2.9~\mu$m and varying wedge angles $\alpha=(15,20,25,30,35,40,55,90)^\circ$. Calculated  electric field norm of the fundamental mode is shown from c) - e) for selected geometries. f) Correspondence between wedge angle and $D_2$ as the cavity radius is varied $R=(80\text{(blue)},100\text{(orange)},140\text{(green)})~$\textmu m. g) The colormap represents $D_2$ as a function of cavity radius and wedge angle. The black dashed line mark the zero GVD points; the curves in (f) correspond to vertical linecuts from (g). Mode and $D_2$ calculations were carried out at $1550$~nm.}
\label{fig:sim}
\end{figure*}

The generation and control of Kerr optical frequency combs in microcavities are fundamental steps towards broadband on-chip coherent light sources~\cite{Gaeta2019}. These sources have been successfully employed in optical communications~\cite{MarinPalomo:2017bv} , frequency metrology~\cite{Papp2014}, spectroscopy~\cite{Dutt2018}, and range measurement~\cite{Suh2018}. Many of these applications often benefit from the dissipative cavity soliton (CS) regime observed  in the anomalous optical group velocity dispersion (GVD) spectral region~\cite{Gaeta2019}. Mid-infrared light sources based on  parametric oscillation also explores dispersion~\cite{Sayson2019} control in the normal GVD regime. Either way, harnessing the properties of microcombs demands for microcavity alternatives with tailorable GVD. Silica wedge microcavities offer a versatile solution for on-chip comb-generation due to its potentially ultra-high optical quality factors ($Q$)~\cite{Lee2012a} and simple fabrication process. In this type of optical microcavity, illustrated in Fig.~\ref{fig:sim}(a), a shallow wedge  is obtained by an isotropic wet etch of the silica glass that not only ensures a smooth glass-air boundary but also reduces the optical-field overlap with the etched glass surface~\cite{Lee2012a}. However, the geometrical effect of the wedge on the GVD is to enhance normal dispersion. As a result, shallow-wedge cavities only exhibit anomalous dispersion at sufficiently large radii, when the wedge-induced GVD is counteracted by silica's anomalous dispersion. Indeed, effective control of zero GVD wavelength in such large radii cavities has been achieved using  two-angle wedges ~\cite{Yang2016} around the 1550~nm telecom band, though it requires a fine control over a nontrivial two-step wet etching.  Closer to  visible wavelengths, effective GVD control  in silica resonators and consequent frequency comb generation  has also been achieved using avoided mode crossing~\cite{lee_towards_2017} or large wedge angles~\cite{li_high-q_2015,Ma2019} achieved using a dry plasma etching.  Although large cavities are advantageous for many frequency comb applications that require GHz-range linespacing~\cite{MarinPalomo:2017bv,Dutt2018}, they usually suffer from small bandwidth due to a universal scaling law ($\Delta f\propto R^{-1}$, $\Delta f$ is the comb bandwidth and $R$ is the cavity radius)~\cite{Coen2013a}. Therefore, smaller cavities with large free spectral range (FSR) are often desirable when targeting at full octave frequency combs~\cite{Okawachi2011a,Pfeiffer2017,Li2017b}.

Here we demonstrate that either anomalous or normal GVD can be achieved at telecom wavelength using small radius (100~\textmu m) silica wedge cavities. Our approach explores strong  photoresist adhesion in a wet etching process that enables unusually large wedge angles. Using this GVD control approach we demonstrate anomalous dispersion frequency comb formation in a broad pump wavelength range. Furthermore, by tuning the tapered fiber bus waveguide diameter we obtain a clean transmission spectrum~\cite{nasir_spectral_2016} that allows for optical mode identification and direct  measurement of GVD. 

%\section{Dispersion control}
Group-velocity dispersion  is strongly influenced by both material and geometric (modal) dispersion. 
In optical cavities with bent sections, the optical angular momentum around curves localize higher frequency waves towards the cavity outer edges, effectively increasing the round-trip time and decreasing the mode FSR, contributing to a normal-GVD. Although such normal-GVD contribution can be overcome by cross-section engineering within ring resonators~\cite{Razzari2009,Pfeiffer2017,Li2017b}, disk cavities offer less degrees of freedom to suppress the normal dispersion contribution. In wedge-shaped microdisk the normal-GVD is further amplified: higher frequency modes (shorter wavelengths) penetrate deeper in the outer wedge tip, further reducing their FSR when compared to longer wavelength modes. Despite silica's anomalous material GVD at telecom wavelengths, this curvature effect is so strong that it can ultimately prevent anomalous GVD within small radii wedge resonators. 

% Although it might seem more intuitive to consider GVD through the wavevector dispersion $\beta(\omega)$,
In optical microcavities it is more meaningful to represent GVD by inspecting how the frequency separation between two adjacent longitudinal optical modes --  the Free Spectral Range (FSR) -- changes along the spectrum. The frequency dependence of the FSR can be expressed by series expansion of optical frequency around a given reference mode,
\begin{equation}
\omega_\mu=\omega_0+\mu D_1 +\frac{1}{2} \mu^2 D_2 +\frac{1}{6} \mu^3 D_3+\ldots,
\label{eq:omega}
\end{equation}
where $\mu=m-m_0$ is an integer label for the mode (relative to a reference mode $m_0$) and $\omega_0$ is a reference mode frequency. As for the coefficients, $D_1/2\pi$ corresponds to the FSR, $D_2/2\pi$ to its variation rate with $\mu$, and so forth.  
Since the FSR of a cavity is given by $\nu_\mathrm{fsr}=v_g/L$ ($v_g$ being the group velocity and $L$ the cavity's length), anomalous GVD ($\partial_\omega v_g>0$) corresponds to a positive $D_2$. 
%one may evaluate its change over one FSR by calculating the variation $D_2/2\pi=\nu_\mathrm{fsr}(\nu_0+v_g/L)-\nu_\mathrm{fsr}(\nu_0)$, which result in the following relation between the usual GVD parameter ($\beta_2=\partial_\omega^2\beta$) and the cavity GVD parameter, $D_2/2\pi=-2\pi v_g^3\beta_2/L^2$. Therefore, anomalous GVD ($\beta_2<0$ or $\partial_\omega v_g>0$) corresponds to a positive $D_2$.

% Using a finite-element simuluation (Comsol Multiphysics), we calculate the possible modes for wavelengths within the range 1500 - 1600 nm. With the aid of the Comsol - Matlab interface, we determined the dispersion  of the first 20 modes for several goemetric configurations of the wedge.

The qualitative arguments given above regarding the GVD are confirmed when we analyze the wedge microdisk cavity through numerical simulations. 
%As depicted in Fig.~\ref{fig:sim}(a), the wedge angle of the microdisk is defined as the arc of tangent given by the ratio between the thickness $h$ of the silica layer and the difference $\delta R \equiv R_\text{out}-R_\text{in}$ between the outer and inner wedge radii. 
As far as mode polarization is concerned, we follow the convention used in \cite{Lee2012a} where the TM (transverse magnetic) mode is the one whose transverse electric field is roughly oriented along the bottom flat oxide-air surface. In the following  discussion we will always be referring to the TE (transverse electric) mode, as it usually has the highest quality factor due to its lower overlap with the wedge surface.
% As previously reported in \cite{Lee2012a}, large radius and thick wedge disks, e.g., 3 mm diameter and $h\approx 8~\mu$m,  are dominated by material dispersion and therefore exhibit overall anomalous dispersion (negative $D_2$) in the C band. Furthermore, the cross-section dielectric guiding effect in further increases the magnitude of the normal $D_2$ of thinner wedge resonators\cite{Lee2012a}. 
%what?
 Figure \ref{fig:sim}(b) presents the simulated residual dispersion, $D_{int}/2\pi= (\omega_\mu-\omega_0-\mu D_1)/(2\pi )$ highlighting the GVD contribution in Eq. \ref{eq:omega} for the fundamental mode of a small radius thin wedge cavity ($2.9~\mu$m thick and $R_\text{out}=100$~\textmu m). as the wedge angles varies from $17^\circ$ ($R_{in}=90$~\textmu m) to $90^\circ$ ($R_{in}=100$~\textmu m). Figures~\ref{fig:sim}(c-e) show the effect of the wedge angle in the total electric field distribution -- the wedge shape can concentrate short-wavelength modes towards its tip. The interplay between the wedge-induced dispersion and cavity curvature is quantified in Fig.\ref{fig:sim}(f): at shorter radii, larger wedge angles are required to achieve zero GVD. 
 %This is expected from our qualitative analysis insight that the cavity's cylindrical nature enhances the degree of normal GVD.
 In Fig.\ref{fig:sim}(g) we show a full map of the GVD (represented by $D_2$) when both radius and wedge angle are varied, the zero-GVD loci confirms that large angles are necessary to balance the curvature-induced GVD contribution.

\begin{figure*}
\centering
  \includegraphics[scale=1]{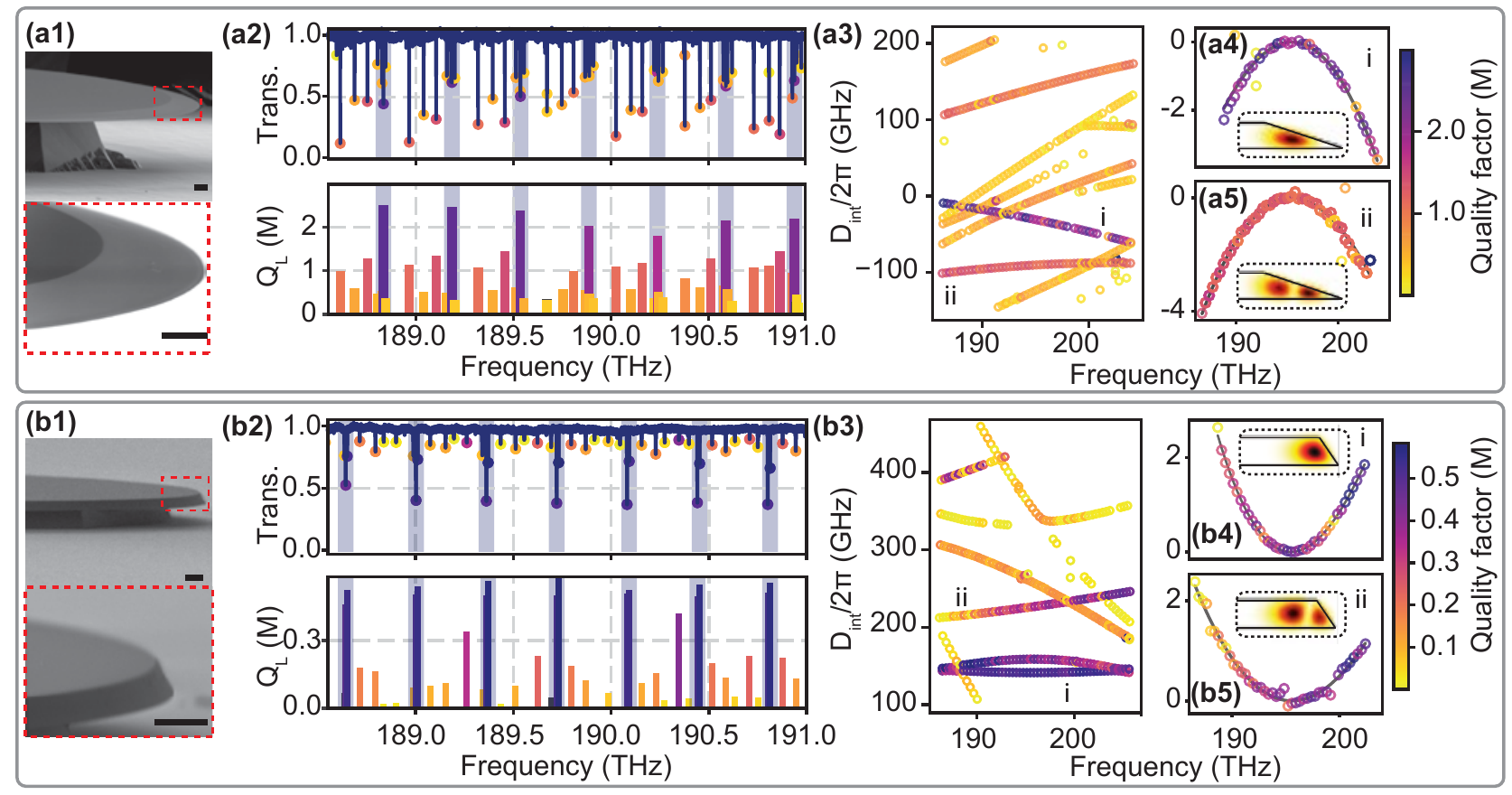}
  \caption{Linear characterization of (a) shallow ($\alpha=17^{\circ}$) and (b) steep ($\alpha=55^{\circ}$) wedge disks. (a1, b1) SEM images of the devices. The bars correspond to $5$~\textmu m. (a2, b2) Cavity transmission (top) and loaded quality factor (bottom) as a function of pump frequency. The colors in the dots correspond to the  $Q_L$ value represented by the colorbar scale. (a3, b3) Measured residual dispersion for the resonances identified in (a2, b2). The modes labeled as \textit{i} and \textit{ii} are shown (a,b)4 - (a,b)5. Eq. \ref{eq:omega} fit yields $D_2/2\pi$ equal to: (a4) $-11.5(7)$~MHz, (a5) $-12.1(6)$~MHz, (b4) $12.0(1)$~MHz, (b5) $6.8(2)$~MHz. The insets show electric field norm.}
  \label{fig:disp}
\end{figure*}
%\cite{long_silica-based_2017}
In order to fabricate both shallow and steep wedge resonators we employed a technique devised in~ \cite{Lee2012a}, starting with a  $2.9$~\textmu m SiO$_2$ layer grown by wet oxidation over a silicon substrate. As previously reported~\cite{Lee2012a}, shallow angles result from the gradual peeling of the resist pads, caused by infiltration of the etchant between the photoresist  and the silica layer. The uniformity of the wedge surface is a consequence of the prolonged isotropic etching, combined with the moderate adhesion of the resist to the substrate. This is what renders achieving large wedge angles a challenging task when using resist soft masks; for instance,  the largest angles reported in~\cite{Lee2012a} are 27$^\circ$, while hard-mask (Cr) process has been used to achieve almost wedge-free disks (although with rather low quality factor)~\cite{long_silica-based_2017}. To ensure a uniform peeling rate, we employed photoresist (Shipley sc-1827) to fabricate the shallow wedges, while large angle resonators were obtained by using MaN-2405 electron resist that has strong adhesion. In both cases, resist adhesion was improved  using surfactant SURPASS 3000 and resist reflow (temperature of $125^{\circ}$C for sc-1827 and $140^{\circ}$C for MaN-2405, both for 5 min). The electron resist was exposed with an electron beam, but it can also be exposed using deep UV light. The silicon undercut was performed using a TMAH (Tetra-Methyl-Amonium Hydroxide) wet etch, which results in an anisotropic pedestal. The fabricated devices are shown in Fig.~\ref{fig:disp}(a1,b1). Using an atomic force microscope, we determine the wedge angles to be approximately $17^{\circ}$ and $55^{\circ}$.

%\section{Dispersion characterization and comb generation}

The GVD and the optical quality factor of the devices were obtained from  the transmission spectra, which were measured by bringing a straight tapered fiber in contact with the microcavity. The transmitted signal from a tunable external cavity laser (Yenista Tunics Reference SCL) was detected with a 1 GHz bandwidth photodetector (New Focus 1623) and recorded by a real-time oscilloscope (Agilent DSO9023A, 2 GHz, 20 Msamples). For the purpose of dispersion measurement, it is of uttermost importance to have a precise frequency calibration for the scanning laser, which was obtained by simultaneously monitoring the transmission of a  fiber-based (Corning SMF28) Mach-Zehnder interferometer (calibrated with $D_1/2\pi$=137(10) MHz, $D_2/2\pi=0.5(1)$ Hz) and an acetylene gas cell. We also verified that the laser scanning speed (10 nm/s) was not limiting our quality factor measurements by comparing them with narrow-band measurements obtained with piezo-scanning of the laser frequency.
%The tapered fiber diameter was carefully optimized  to reduce excitation of higher order cavity modes, which cleans the transmission spectrum and facilitates the identification of the mode families,  as exemplified by figures \ref{fig:disp}(a2) and (b2). 
The fundamental optical mode is thus isolated by employing thicker tapers ($\approx 3.3$ \textmu m) that decrease the phase-matching between the  taper optical mode and the higher order disk modes \cite{nasir_spectral_2016}. Using the full bandwidth of our laser (1460 - 1610 nm), we measured the fundamental TE mode loaded quality factor ($Q_L$) to be around $6\times10^5$ for the steep wedges, and $2.2\times10^6$ for the shallow wedges, as shown in Figs. \ref{fig:disp}(a2) and \ref{fig:disp}(b2). Scanning electron microscope (SEM) images of the devices (shown in figures \ref{fig:disp}(a1) and (b1)) suggest that this discrepancy is due to lithography-related sidewall roughness rather than the actual wedge angles. 
%Other wedge cavities  fabricated by wet etching methods~\cite{Lee:2012hn} indicate that it should be possible to achieve $Q\approx 10^7$ for disks with  similar thicknesses.
\begin{figure*}
\centering
  \includegraphics[scale=1]{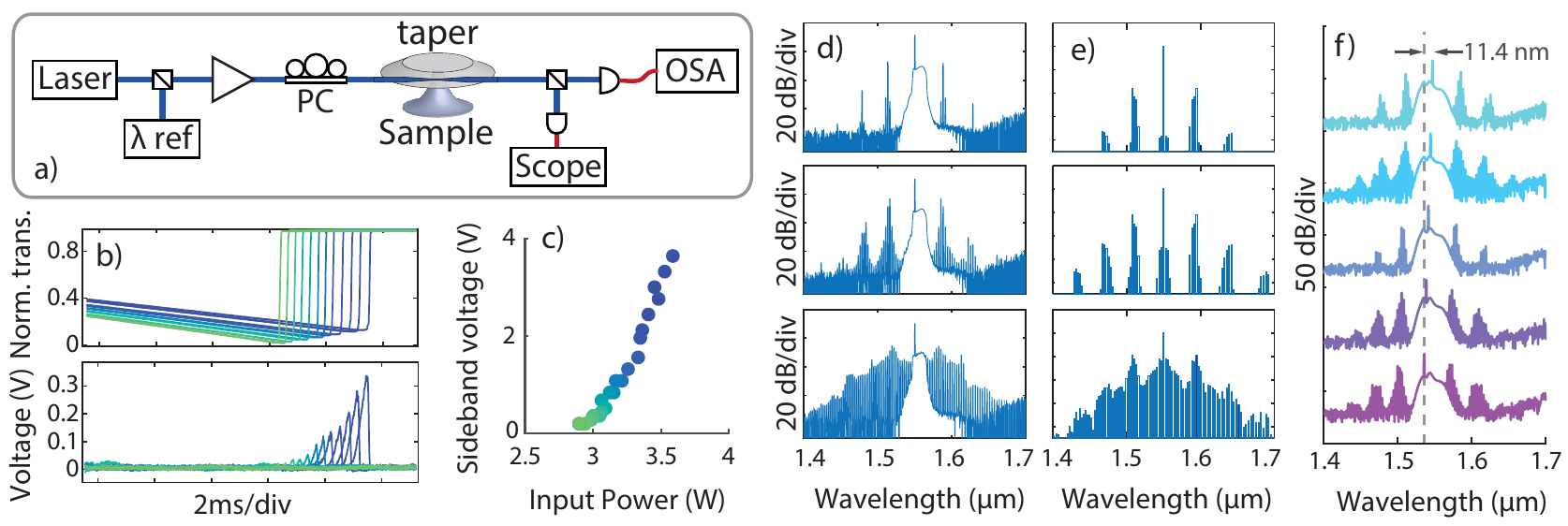}
  \caption{(a) Experimental setup for Kerr comb generation. The CW (continuous-wave) laser is modulated and amplified before the tapered fiber. $\lambda$ ref: Mach-Zehnder interferometer and acetylene cell for wavelength reference. PC: polarization controller. OSA: optical spectrum analyzer. (b) Top: transmission spectra of a resonance at $\approx 1547$ nm. Bottom: First sidebands being generated near minimum transmission points. The input power increases from green to blue. (c) Sideband intensity versus input peak power.  (d) Experimental and (e) simulated formation dynamics of the comb at 1547.11 nm. The power increases from top to bottom. (f) Comb spectra obtained with the OSA for varying pump wavelengths. The dashed line marks the pump wavelength for the first comb, 1535.73 nm. The combs are generated at every FSR$=2.85$ nm.}
  \label{fig:comb}
\end{figure*}

A general picture of the GVD in our devices can be obtained by inspecting Fig.~\ref{fig:disp}~(a3, b3). These curves are obtained in three steps: (i) we assign a single relative azimuthal number to all modes lying within one FSR in the measured transmission, Figs. \ref{fig:disp}( a2, b2). Their resonant frequencies are then (ii) plotted against the azimuthal numbers; a straight line fitted to this data corresponds to an average FSR ($\overline{D}_1$) of the measured cavity modes. Next, (iii) an overall picture of the GVD ($D_2$) for all modes around $\omega_0$ is obtained by subtracting ($\omega_{\mu}-\omega_0-\overline{D}_1\mu$) from the dataset, where $\mu=0$ corresponds to the inspection frequency $\omega_0$. For example, in Figs. \ref{fig:disp}(a3, b3) the chosen inspection frequency is $\omega_0/2\pi=195.3$~THz ($\lambda=1535$~nm) and the subtracted linear trend corresponds to an average FSRs of  $\overline{D}_1/2\pi\approx350$~GHz ($\sim 2.8$~nm).  The significant difference in slopes noticeable in Figs. \ref{fig:disp}(a3, b3) reflects the distinct FSRs of each modal family. Avoided mode crossings within higher order modes can be easily identified in this representation. From Figs. \ref{fig:disp}(a3, b3) we can slice specific optical mode branches from the dataset and fit Eq.~\ref{eq:omega} to them. In this last step, the precise values for $D_1,D_2$ of the chosen branch are calculated
%; for instance $D_1/2\pi=350.64(1)$ GHz  (Fig.~\ref{fig:disp}(a4)) and $D_1/2\pi=351.927(8)$ GHz (Fig.~\ref{fig:disp}(a5)). 
In Figs.~\ref{fig:disp}(a4, a5, b4, b5) we show the  fitted model of Eq.~\ref{eq:omega}  for two modes mode branches with highest $Q_L$ of each device. As expected from Eq. \ref{eq:omega}, the residual dispersion $D_\text{int}$ for each mode family appear as parabola-shaped curves centered at the inspection frequency. In the small-angle wedge ($\alpha=17^\circ$) of Fig.~\ref{fig:disp}(a4, a5) we obtain normal GVD;
%$D_2/2\pi=-11.5(7)$~MHz and $D_2/2\pi=-12.1(6)$~MHz
whereas  anomalous GVD is obtained for the large-angle wedge.
%($D_2/2\pi=12.0(1)$~MHz and $D_2/2\pi=6.8(2)$~MHz, respectively).
We identify these modes as the first two lowest order optical modes by comparing their measured $D_1$ and $D_2$) with numerical simulations. 

%  This translates to the concavity of the parabola shown in Fig.  \ref{fig:disp}e), $D_2/2\pi=12.7$~MHz and $D_3/2\pi=-78$~KHz,  from which we determine the dispersion to be anomalous \cite{herr_temporal_2014}. The measured values are summarized in Table \ref{tab:values}.
 
% \begin{table}[htbp]
% \centering
% \caption{\bf Characterized values}
% \begin{tabular}{c||c|c}
% \hline
% Resist & sc-1827 & MaN-2405 \\\hline
% $\alpha$ & $17^{\circ}$ & $55^{\circ}$ \\
% $Q$ & $0.9\times 10^6$ & $0.3\times 10^6$ \\
% $D_2/2\pi$ & -9.7~MHz & 12.7~MHz \\
% $D_3/2\pi$ & ~KHz & ~KHz \\
% \hline
% \end{tabular}
%   \label{tab:values}
% \end{table}

%\section{Frequency comb generation}

Using the anomalous GVD engineered resonator,  we were able to consistently generate optical frequency combs by pumping the resonator at the optical mode shown in Fig.~\ref{fig:disp}(b4). Figure \ref{fig:comb}(a) outlines the experimental setup used for comb generation: the amplitude of the CW Tunics laser is modulated by an EOM (electro-optical modulator) and then amplified with an EDFA (erbium doped fiber amplifier), concentrating up to 14 W (peak power) into 40 ns pulses (1 MHz repetition rate). The pulse duration is long enough to ensure a quasi-CW regime. The cavity transmission is analyzed with an oscilloscope and an OSA (optical spectrum analyzer).

Choosing a resonance at approximately 1550 nm, we determined the threshold power for comb generation and studied its formation dynamics. The threshold was measured by simultaneously monitoring the pump transmission and the power of the first generated sideband while scanning the laser frequency with a piezo-actuator. The sideband power was measured by the OSA centered at 1527 nm with zero span and resolution bandwidth of 2 nm, resulting a voltage signal proportional to its power. The results are presented in Fig.~\ref{fig:comb}(b), where the color scale shifts from green to blue as input power increases. We notice that the first comb sidebands are generated towards the end of the bistable transmission and its intensity diminishes as the resonance minimum is blue-shifted at smaller input powers. The sideband maximum power evolution (Fig.~\ref{fig:comb}(c)), allows us to identify the threshold near 2.9~W, which is in reasonable agreement with the predicted threshold for comb generation (2 W for an effective mode area $A_\text{eff}=5$~\textmu m$^2$) )~\cite{Herr2012b}. 
% Although this value is rather high, it could be decreased by a factor up to $10^4$ if our quality factor improves to match the previously reported in the literature for wedge resonators~\cite{Lee2012a,li_high-q_2015}.

For the comb formation measurements, we captured the OSA spectra as the laser (14 W input peak power) is tuned towards the bistable resonance in 1~pm steps. Figure~ \ref{fig:comb}(d) highlights the different formation stages: at first, primary sidebands are generated 13 FSRs away from each other. Then, secondary sidebands rise around the primary lines in adjacent modes, until the sub-combs merge into a single broader comb, as expected for small FSR devices~\cite{Herr2012b}. The coupled mode equations for the entire system were solved with the open-source CombGUI toolbox, and the simulation results based on the measured experimental parameters are in good agreement with the experimental data, as shown in Fig.~\ref{fig:comb}e). 
% The simulations also indicate the possibility of soliton formation, although we did not attempt to experimentally access the red-detuned region where solitons are known to be excited.
Finally, since our modal dispersion engineering does not rely on avoided crossings~\cite{Kim2017}, combs could be excited at every FSR step of the pump laser. Fig.~\ref{fig:comb}(f) shows some example combs generated within an 11 nm range ($\sim$4 FSRs) around 1550 nm.

%\section{Conclusions}
 
In conclusion, we demonstrate that small radii wedge resonators can be engineered to enable either normal or anomalous GVD. Since our approach does not rely on avoided crossings, not only it allows for comb excitation in a broad range of pump frequencies but also may be more easily explored towards octave spanning frequency combs. Our GVD engineering strategy, if combined with disk thickness control, could also be extended towards other frequency ranges or wedge materials~\cite{Ramiro-Manzano2012,Kang2017} . Extension of our results may allow the control of higher order GVD (e.g. $D_4$) and enable phase-matching at exotic wavelengths.
% By achieving large wedge angles using wet chemistry etching, we may expect to achieve ultra-high quality factors by optimizing the lithography process, and thus reducing the required pump power for comb generation. Although we did not focus on intermediate wedge-angle, this can be achieved by adjusting the resist adhesion and may allow controlling of higher order GVD (e.g. $D_4$) to phase-matching at exotic wavelengths.

\noindent \textbf{Funding}: this work was funded by FAPESP (2012/17610-3, 2012/17765-7, 2016/05038-4, 2018/15577-5, 2018/21311-8, 2018/15580-6, 2018/25339-4), Coordenação de Aperfeiçoamento de Pessoal de Nível Superior - Brasil (CAPES) (Finance Code 001), and CNPq. \textbf{Acknowledgements}: we acknowledge Antônio Von Zuben for his support in the microfabrication, and Center for Semiconductor Components and Nanotechnologies and MultiUser Laboratory (LAMULT) for providing the micro-fabrication infrastructure. \textbf{Disclosure}: the authors declare no conflicts of interest.

% Bibliography
\bibliography{references}
%\bibliographyfullrefs{references}

\end{document}